\documentclass[final,5p,times,twocolumn]{elsarticle}

\usepackage{amsmath}
\usepackage{amssymb}
\usepackage{amsfonts}
\usepackage{graphicx}
\usepackage{dcolumn}
\usepackage{fancyhdr}
\usepackage{bm}

\journal{Cond. Matt.}

\begin{document}

\begin{frontmatter}



\title{Vitrification, relaxation and free volume in glycerol-water binary liquid mixture: Spin probe ESR studies}


\author{Debamalya Banerjee}
\author{S. V. Bhat \corref{1}}
\ead {svbhat@physics.iisc.ernet.in}
\cortext[1]{Corresponding author}
\address{Department of Physics, Indian Institute of Science, Bangalore 560 012, India}

\begin{abstract}
Glass transition and relaxation of the glycerol-water binary mixture system are studied over the glycerol concentration range of 5 - 85 mol\% using the highly sensitive technique of spin probe ESR. For the water rich mixture the glass transition, sensed by the spin probe, arises from the vitrified mesoscopic portion of the binary system. The concentration dependence of the glass transition temperature manifests a closely related molecular level cooperativity in the system. A drastic change in the mesoscopic structure of the system at the critical concentration of 40 mol\% is confirmed by an estimation of the spin probe effective volume in a temperature range where the tracer reorientation is strongly coupled to the system dynamics. 

\end{abstract}

\begin{keyword}
\sep {Glass transition, Free volume, Electron Spin Resonance, Phase Separation}
\PACS {61.43.-j, 64.70.P-, 76.30.-v}

\end{keyword}

\end{frontmatter}


\section{Introduction}
Thermodynamic, transport and relaxation properties of binary liquid mixtures have been the subject of scientific investigation for the last few decades. The presence of two different masses and length scales of interactions is largely responsible for the composition dependence - often non-linear as suggested by experiments \cite{Indraswati:2001,Tomasik:1990} and theoretical studies \cite{Saksena:1975,Bagchi:2001} - of static and dynamic properties of such systems. Nevertheless, the tunability offered by this composition dependence of certain properties is of great use in both industrial and laboratory chemistry and biology.

The glass transition temperatures of a binary mixture system is usually a monotonic function of the composition and can be well described by the empirical equation of the form \cite{Gordon:1977}
\begin{equation}
\label{emp}
T_g = w_1\, T_{g_1} + w_2\, T_{g_2} + K\, w_1 w_2
\end{equation}
where '$w$'\;s represent the weight fractions, '$T_{g}$'\,s are the glass transition temperatures of the two components respectively and $K$ is a constant. Aqueous solutions also follow this concentration dependence of $T_g$ with a few exceptions \cite{Murthy:2008}.

{For extensively hydrogen bonded glycerol-water (G-W) mixture, when the glycerol concentration is above 40 mol\%, water molecules associate nicely into the H-bonded network of glycerol. In this range of concentration the G-W mixture vitrifies completely on quenching \cite{Hayashi:2005, Inaba:2007}. The number of hydrogen bonds associated with glycerol is $n_G \; \simeq$ 6 \cite{pedro:97} compared to $n_W$ = 4 for water. So, for a 40 mol\% glycerol concentration of a G-W mixture, all the glycerol hydrogen bonds can be occupied by water molecules and so, this concentration is often referred as the 'critical concentration'\cite{Hayashi:2005a}. However, below $40$ mol\% of glycerol concentration mesoscopic inhomogeneities in the form of cooperative domains of water, where molecular reorientation occurs cooperatively by means of dipolar coupling, are found in the G-W mixture \cite{sudo:2002, Hayashi:2006}. This picture was first proposed by McDuffie {\it et al.} \cite{McDuffie:1962} and confirmed by recent investigations carried out using adiabatic calorimetry \cite{Inaba:2007}, Differential Scanning Calorimetry (DSC) \cite{Hayashi:2005a} and Broadband Dielectric Spectroscopy (BDS) \cite{sudo:2002,Hayashi:2006}. Lowering of the glycerol content from the critical value of $40$ mol\% results in the growth of these water cooperative domains and for a glycerol concentration $<20$ mol\%, the formation of ice nanocrystal \cite{Hayashi:2005a,sudo:2002} along with interfacial water layer separating the former phase from the vitrified portion at low temperatures was reported \cite{Hayashi:2005a}. A water exchange kinetic equilibrium was proposed for this range of glycerol concentration between the vitrifies portion of G-W matrix and the interfacial water layer which reportedly keep the concentration of the mesoscopic portion almost constant to a value close to the critical concentration \cite{Hayashi:2005a}. Consequently, the T$_g$ variation of the system with concentration was seen to attain a plateau for lower glycerol contents \cite{Hayashi:2005a,Inaba:2007}.}

{Molecular probes are well known for the study of cooperativety and rotational diffusion of glassformers \cite{marcus:95,AndreozziEtAl:1996a,Zondervan:2007}. In a system like glycerol-water which is intrinsically inhomogeneous for lower glycerol content, molecular probing can provide useful information about cooperativity and local structures. This motivates us to study the G-W system using the molecular probe technique of spin probe ESR. }

Electron paramagnetic resonance or electron spin resonance (EPR/ESR) measures the energy level separation of unpaired electrons. This technique is well known for its high sensitivity. To investigate (mainly the relaxation and glass transition) the systems having no intrinsic ESR sensitive center with this technique, the method of spin probe ESR is widely used. In this method a small amount of a spin probe (also called a tracer) which is generally an organic free radical with one ~{\bf N-O} group, is added in small amount to the sample under study. The detected signal is the resonant microwave power absorption spectrum of the unpaired electron of the ~{\bf N-O} group when exposed to an external magnetic field. This unpaired electron is coupled to the nuclear spin of the nitrogen molecule through hyperfine coupling which causes the spectrum to symmetrically split into three lines in a isotropic medium \cite{Berliner:1976}. The local environment of the host matrix has significant effect on the re-orientation mechanism of the guest tracer molecules to such an extent that the tumbling frequency of the tracer is found to be closely associated with the structural relaxation process of the system \cite{Kumler:1976,AndreozziEtAl:1996a}. Experiments can be performed spanning a relaxation timescale of $\sim 10^{-7} - 10^{-12} s$ typically in the temperature range of 4.2 - 300 K. Dynamics and glass transition of different polymers and model glassformers have been studied using the technique of spin probe ESR \cite{Boyer:1980,Andreozzi:2006a}. This technique has been successfully used for $T_{g}$ determination of PEG-2000, where the glass transition lies beyond the detectability of conventional DSC attributed to the high degree of crystallinity of the polymer \cite{Bhat:2004}. Amorphous solid water (ASW) is another example where the glass transition and relaxation have been studied by the method of spin probe ESR and the glass transition temperature, which could not be measured directly by conventional techniques \cite{Angell:2002}, is detected at $\sim 132 K$ \cite{Bhat:2005}, close to the putative glass transition temperature of $136 K$.

The glassformers are commonly classified based on their degree of fragility closely related to its structural relaxation process above T$_{g}$, where amorphous solid turns into liquid. The structural relaxation time $\tau_s$ (alternatively the shear viscosity $\eta$) of this supercooled liquid often follows a non-Arrhenius temperature dependence which is well described by the VTF equation of the form
\begin {equation}
ln \frac {\tau_s}{\tau_{0}}= \frac{B\;T_{0}}{T-T_{0}} 
\label{VTF}
\end {equation}
with $\tau_0 = \tau_{s\,\vert T \rightarrow \infty}$, $B_0$ and $T_0$ are constants. When $ln \; \tau_s$ (or $\eta$) is plotted against inverse temperature, {\it i.e.} the so called 'Arrhenius plot', the  curvature is a measure of the fragility which can be further quantified by the fragility index '$m$', equal to the slope of this plot at $T = T_g$ \cite{Angell:1993}. In this respect glycerol is an 'intermediate' glassformer \cite{Angell:1993} whereas weakly supercooled water is fragile by nature \cite{ItoAngell:1999}. So, a steady decrease in fragility of the G-W system with increasing glycerol concentration is expected as observed in the recent BDS studies \cite{sudo:2002,Hayashi:2006}. 

  The rotational diffusion of a spheroid tracer molecule in a fluid is well accounted by the Debye-Stokes-Einstein (DSE) law of the form \cite{Blanchard:1997}
\begin {equation}
\tau = \frac{V \eta}{k_{b}T}\; f \: C_{bc} 
\label{DSE}
\end{equation}
 where $\tau$ is the rotational correlation time, $\eta$ is the shear viscosity, $T$ and $k_{b}$ are the temperature and Boltzmann constant respectively. $V$ is molecular effective volume, $f (\ge 1)$ is the shape factor related to the non-spherical shape of the tracer particle and the slip factor, $C_{bc}$ accounts for the variable friction under different boundary conditions ($C_{bc} = 1$ for {\it stick limit} \cite{Laia:2002}). The expression for $C_{bc}$ under slip boundary condition is well known \cite{Zwanzig:1974} and the shape factor is a function of the ratio of axial dimensions $\rho$ \cite{Blanchard:1997}. The DSE relation is well validated for spin probe dissolved in a liquid or a weakly supercooled host. But a significant deviation is observed at low temperatures, particularly in the vicinity of glass transition temperature \cite{AndreozziEtAl:1997,AndreozziEtAl:1996a}.

In this paper we demonstrate the glycerol concentration dependence of glass transition temperature for G-W binary mixture system. For glycerol content $> 40$ mol \% we have found that the T$_{g}$ variation is well described by the empirical relation given by equation \ref{emp}. For  $< 40$ mol \% composition where the $T_{g}$ values obtained by any technique has only limited meaning due to the well established phase segregation in this range, our studies throw light on the spin probe re-orientation processes in the supercooled host matrix in the temperature range of 230 - 305 K across the concentration range with particular emphasis on the free volume of the system.

\section{Experimental}
Eight samples have been prepared within a range of 5 - 85 mol\% of glycerol concentration using as obtained anhydrous glycerol from Sigma Aldrich (Mol. Wt. 92.1, 99\%, GR grade) and triple distilled, de-ionized water. Almost spherical organic tracer molecule TEMPO (2, 2, 6, 6-tetramethyl-l-piperidine-l-oxy, Mol. Wt. 156.24) was added and evenly mixed to the sample by a magnetic stirrer arrangement (typical spin probe concentration $\sim 10^{-3}$ mol/lit). Samples were quenched by rapid cooling down to 4.2 K in a pre-cooled Oxford continuous flow liquid helium cryostat with ITC 503 temperature controller attached to a Bruker EMX X-band ESR spectrometer. The cooling protocol offers an uncontrolled but high cooling rate $\geq$ 200 K/min. Spectra were recorded, as usual in the derivative mode with lock in detection by sweeping the magnetic field, on heating in the temperature range of 130 - 305 K. At any selected temperature, sufficient waiting time was allowed for stabilization until no aging, {\it i.e.} spectrum evolution was detected.

\section{Results and discussion}
\subsection{Vitrification}
\begin{figure}[t]
\includegraphics [width = 9 cm]{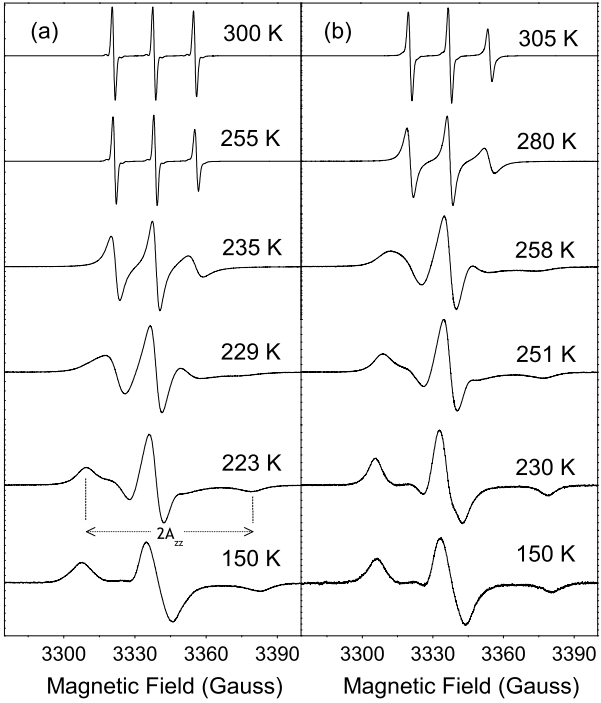}
\caption{\label{fig1}
Temperature variation ESR spectrum for a) 11 mol\% and b) 69 mol\% glycerol. Tracers are trapped in disordered host at low temperature and give rise to the 'powder pattern'. The high temperature triplet spectrum is the result of motional averaging of magnetic anisotropies in a liquid host. The extrema separation (2A$_{zz}$) is marked on the second spectrum from the bottom of left panel}
\end{figure}

We first present the temperature variation ESR spectra for two selected concentrations in fig \ref{fig1}. 
At lower temperatures, the re-orientation of TEMPO molecules is slow and the ESR spectra are almost temperature independent. This rigid limit spectrum, or the so-called powder pattern, lacks any dynamical information and represents the distribution of magnetic anisotropies of Zeeman and hyperfine tensors due to isotropic angular distribution of TEMPO molecules in the disordered, solid host matrix. At higher temperatures the tumbling frequency increases as the consequence of more available thermal energy leading to more dynamic environment of the host and the rotational correlation time enters the ESR detectable time window of  $\tau_{ESR}\sim 5\times 10^{-7}s$. At this point, the molecular axes of TEMPO can sample the whole range of solid angle within $\tau_{ESR}$ and the spectrum unfreezes. 
\begin{figure}[t]
\includegraphics [width = 9 cm]{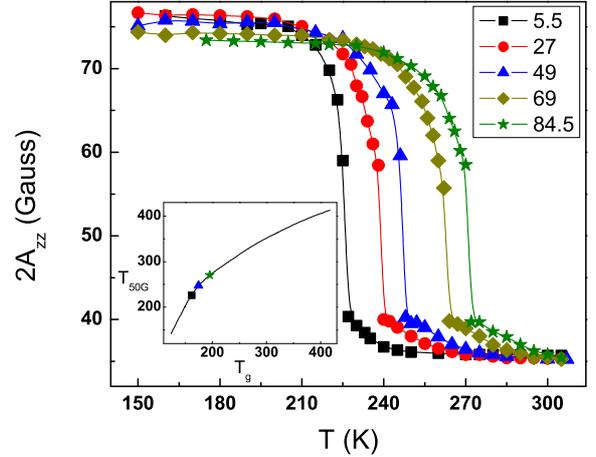}
\caption{\label{fig2}
Temperature variation of the extrema separation 2A$_{zz}$ for selected concentrations. The solid lines are drawn through the data points for clarity. {\bf Inset:} the 'Kumler plot' establishing the correlation between the crossover temperature and the T$_{g}$ \cite{Kumler:1976} with three points from the present work on it. }
\end{figure}
The subsequent averaging of the magnetic anisotropies leads to the narrowing of the overall spectrum (motional narrowing) and the extrema separation of the spectrum, which corresponds to twice the z-principle component of hyperfine tensor $(2A_{zz})$, decreases from its rigid limit value. At even higher temperature, when the tumbling frequency of the probe in a more dynamic host is comparable to the strength of the hyperfine interaction $\sim 10^{8}$ Hz, narrowing is extreme and the lineshape collapses into a symmetrically spit three-lined spectrum of isotropic limit. As a result, the extrema separation reduces dramatically and at a much higher temperature i.e. well inside the Redfield limit, it assumes a value twice the isotropic limit hyperfine tensor $(2A_{0})$. Figure \ref{fig2} represents the temperature variations of the extrema separations for the selected concentrations. A gradual increase in the crossover temperature with increasing glycerol concentration is noticed. The crossover temperature, generally referred as $T_{50G}$, is found to be empirically correlated to the glass transition temperature ($T_{g}$) by the so called 'Kumler plot' \cite{Kumler:1976}(Figure 2, inset). $T_{g}$s as determined using the crossover temperatures by this method are plotted in Fig. \ref{fig3} as a function of glycerol concentration. Slightly higher T$_g$ values obtained compared to the literature can be attributed to the high cooling rate as predicted theoretically \cite{Sastry:1998}. Three different regions which are observed in the T$_g$ variation with respect to glycerol concentration (fig \ref{fig3}), have been distinguished as region I, II and III. In region I, the glass transition temperature is only a weak function of concentration where region II is characterized by stronger variation. Region III, where the components are homogeneously mixed \cite{Hayashi:2006}, exhibits the strongest variation among all and it is well fitted to equation \ref{emp} with $T_{g} = 132\, w_{H20} + 200.5\, w_G -95\, w_G w_{H20}$ (Fig \ref{fig3}, inset). Here we recall the glass transition temperature of bulk water is estimated $\sim$132 K by the same method \cite{Bhat:2005}. The crossovers between the regions of $T_g$ variation are marked by vertical arrows.

The observed deviation of T$_g$ dependence from the trend followed by the homogeneous region III for $<$ 40 mol\% glycerol contents can be taken as the evidence of the development of inhomogeneities. Recent BDS studies have identified such inhomogeneities as the water cooperative domains \cite{sudo:2002,Hayashi:2005a}. At room temperature, spin probe molecules have full access to these domains formed inside a liquid host and information to the extent of mesoscopic scale can be extracted from the spectra which will be discussed later. If the size of the water cooperative domains are not small enough to avoid crystallization, the domains freeze to form ice nanocrystals on quenching as reported in the BDS studies for a glycerol concentration $<$ 20 mol\% \cite{sudo:2002,Hayashi:2005a}. The spin probe molecules would occur as impurities inside the ice crystals and thus will be disposed out into disordered neighboring environment of vitrified mesoscopic G-W matrix. Here we recall that crystallization is one of the most efficient method of purification. For a glycerol concentration between 20 and 40 mol\%, the smaller water cooperative domains may vitrify on quenching to form droplets of amorphous solid water (ASW). On heating, the ASW turn into ultraviscous liquid above its glass transition temperature $\sim$136K and crystallize into cubic ice $\sim$150K \cite{Stanley:1998,ItoAngell:1999}. So the spin probes trapped inside these vitrified ASW domains, as described above, will eventually be disposed into the disordered neighboring environment at a temperature $\sim$150K. 
\begin{figure}[t]
\includegraphics [width = 9 cm]{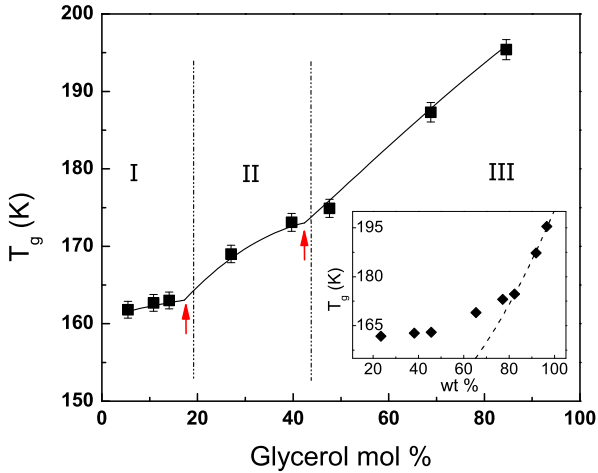}
\caption{\label{fig3}
Glass transition temperature T$_{g}$ as a function of mole percentage for G-W mixture. The crossover between different regions are marked by vertical arrows. {\bf Inset:} The T$_{g}$s as a function of corresponding weight percentages. Dotted line is the fit to the empirical equation \ref{emp}.}
\end{figure}
We further substantiate our conclusion about the spin probe environment with following arguments: The entrapment of spin probe molecules in ice is supposed to give well known freezing signature in an ESR study \cite{Ross:1965,LeighEtAl:1971} which is not seen in the present set of experiments. Moreover, entrapment of the spin probe in the ice during quenching will lead to almost temperature independent signal up to the water melting temperature $T_{m}(= 273.15 K)$, where sudden change in the signal shape is expected. Instead, we observe a gradual change in the spectrum with increasing temperature and the spectrum of $\sim$255 K for all the lower concentrations ($<$ 40 mol\%) qualitatively similar to that of the liquid state spectrum of 300 K (fig \ref{fig1}a), hardly understood if ({\it even fraction of}) the spin probe is trapped inside ice. So, we can conclude that no spin probe is trapped in the ice fraction of the frozen matrix and the glass transitions sensed by spin probes is entirely due to the vitrified portion of G-W mixture. 

As the formation of mesoscopic inhomogeneities force the T$_g$ variation below 40 mol\% to follow a different course from that of the homogeneous region III, DSC and adiabatic calorimetry studies suggested the concentration of the vitrified portion of G-W mixture remains close to the critical concentration of 40 mol\% \cite{Hayashi:2005a,Inaba:2007}. Therefore the T$_g$\,s of this region should remain close to that of the 40 mol\% G-W mixture. But our result shows considerable lowering of $T_{g}$\,s from that of the 40 mol\% value in region II, though the lowering is not as much as it would have been following the concentration dependence of homogeneous regime. For a glycerol concentration below the critical value of 40 mol\%, the configuration where all the hydrogen bonds of glycerol is occupied by water molecules with excess water molecules localizing cooperatively is a configuration energy minimum. But well inside the liquid regime, a system is known to exhibits large scale thermal fluctuations by sampling the higher energy configurations with a broad distribution of configuration energy around an average value \cite{Stillinger:1995,Sastry:1998}. The high energy configurations in the present case can be identified with a state where excess water is not only confined within the cooperative region but to some extent diffused into the region of the glycerol-water interactions, thus lowering the concentration of the latter phase from its critical value. The high cooling rate used in the present set of experiment does not allow the system to achieve a low energy configuration by molecular rearrangement during the thermal quench. Instead, the frozen state of low temperature largely retains the configuration of liquid state before quenching. So, the spin probes most likely get trapped into a vitrified portion of G-W matrix having a concentration lower than the critical concentration. This explains our finding on the significant deviation of T$_g$ s for lower glycerol content from that of the critical composition value. Moreover, the presence of another distinct region where T$_g$ variation with concentration is negligible (region I) suggests presence of additional inhomogeneities at lower glycerol concentrations. Such inhomogeneities may be identified with the interfacial water layer formed between the ice nanocrystals and vitrified G-W mixture below 20 mol\% of glycerol content during thermal quench as observed in BDS study \cite{Hayashi:2005a}. For a G-W mixture in this concentration range, the glycerol-water proportion of the mesoscopic region might stay close to a constant value by means of a proposed water exchange kinetics with the interfacial water layer \cite{Hayashi:2005}. For a mixture with glycerol concentration lying between 20 mol\% and 40 mol\%, interfacial water layer does not form, resulting a stronger variation of T$_g$ with concentration is seen in region II. This argument is further supported by the fact that the crossovers between different regions of the concentration dependence of observed T$_g$ occur close to the glycerol concentrations $\sim$40 mol\% and $\sim$20 mol\% (fig 3), marked by BDS for the occurrence of water cooperative domains and interfacial water layer at low temperatures \cite{Hayashi:2005a}, respectively.

\subsection{Relaxation}
As discussed before, the spin probe reorientation dynamics is heavily influenced by the host matrix environment. If the host is a glassformer, the environment becomes more mobile above its glass transition temperature. A recent study on poly(propylene glycol) shows that the free volume of the glassformer and the tracer reorientation in it are strongly coupled \cite{Andreozzi:2006}. In the vicinity of $T_{g}$, the free volume of the host remains smaller than the intrinsic volume of the tracer molecule and the probe reorientation takes place in a tight, almost solid like environment. In this range, the temperature dependence of $\tau$ is well described by Arrhenius type relaxation process \cite{Andreozzi:2006, AndreozziEtAl:1996a}. The free volume of the host increases with temperature and at a temperature slightly higher than $T_{50G}$ it becomes comparable, or even larger, than that of the intrinsic volume of the tracer molecule. The free rotation of the tracer in this temperature regime is reflected in its non-Arrhenius relaxation usually represented by a VTF type temperature dependence of equation \ref{VTF} \cite{Andreozzi:2006}. For small molecular glassformer OTP and salol doped with TEMPO, the DSE law is found to be obeyed in this high temperature non-Arrhenius region \cite{AndreozziEtAl:1996a, Andreozzi:2006a}.

Above the crossover temperature $T_{50G}$ of the $2A_{zz}$ transition, the spin probe tumbling is fast enough to average out all the anisotropies associated with the Zeeman and the hyperfine tensor and the lineshape collapse into a narrow, symmetric triplet. The correlation time of this fast rotation can be well estimated using the formula given by Freed and Frankel \cite{Freed:1963}
\begin {equation}
\tau = \frac{\left( \sqrt{\dfrac{h_{+1}}{h_{-1}}}-1 \right) \Delta H }{1.2 \times 10^{10}}\;sec
\label{freed}
\end {equation}
where $h_{+1}$, $h_{-1}$ and $\Delta H_{+1}$ are the heights of the first and the third line and the width of the first line of the narrow triplet signal respectively. 
\begin{figure}[t]
\includegraphics [width = 9 cm]{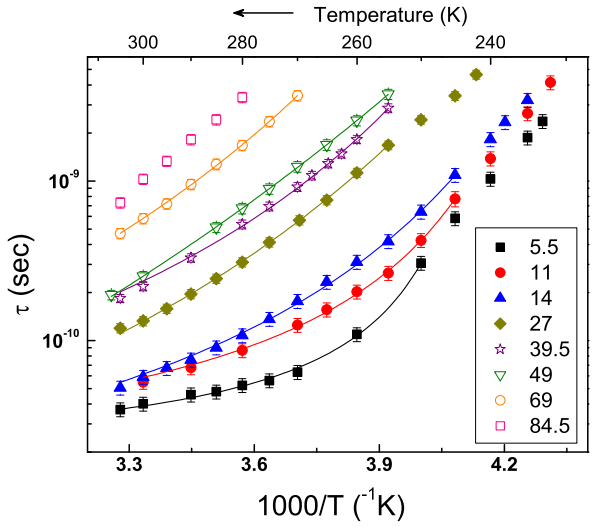}
\caption{\label{fig4}
Spin probe correlation times for different concentration obtained by equation \ref{freed}. Solid lines are fit to the VTF law (equation \ref{VTF}). Note the steady decrease (increase) in fragility with higher glycerol (water) content.}
\end{figure}

The correlation times calculated using equation \ref{freed} for different concentrations are plotted in figure \ref{fig4} as a function of inverse temperature (Arrhenius plot). Solid lines represent the VTF fit to data for the full range of concentration except for 85 mol\%, which could not be fitted because of limited numbers of data points.  A steady decrease in the curvature, which is closely related to the system fragility, is observed from water rich to glycerol rich mixture. The low temperature points for the lower glycerol concentrations, which could not be fitted with VTF law, belong to a regime where the spin probe re-orientation becomes weakly coupled to the viscosity which is attributed to the higher degree of heterogeneity developed in the glass former and the DSE law takes a fractional form \cite{AndreozziEtAl:1996a,AndreozziEtAl:1997}.

\subsection{Free volume}
In a temperature region where tracer rotation obeys the DSE law, the correlation times measured at an isotherm (would be a vertical line in case of fig \ref{fig4} if drawn) can be used along with the available viscosity data \cite{segur:1951} to calculate $V$ using equation \ref{DSE} for a given concentration. In the present case it is the tumbling of the tracer molecule about $r_{\perp}$ sensed by ESR. For a prolate tracer molecule, $f$ takes the form \cite{Blanchard:1997,Laia:2002}
\begin {equation}
f = \frac{2}{3}\frac{1- \rho^{4}}{\left [(2- \rho^{2})\dfrac{\rho^{2}}{\sqrt{1-\rho^{2}}}\ln\left(\dfrac {1+\sqrt{1-\rho^2}}{\rho} \right)\right ] - \rho^2}
\label{slipBC}
\end {equation}
where $\rho$ ($= a/b <1$, $b$ is the axis of the rotation) is the ratio of axial dimensions. By using values of axial radii of TEMPO from reference \cite{AndreozziEtAl:1997}, we find $\rho \approx 0.73$, thus $f \approx 1.13$. The slip factor $C_{bc}$ is estimated to be $\approx 0.07$ for this molecule in the {\it slip limit} \cite{Zwanzig:1974}.

\begin{figure}[t]
\includegraphics [width = 9 cm]{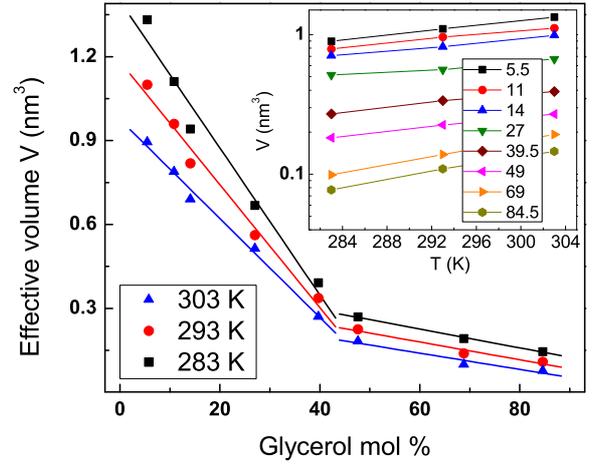}
\caption{\label{fig5}
Effective volume of the probe molecules at different temperatures as a function of glycerol concentration. Note i) the change occurring in $V$ around critical concentration of 40 mol\% ii) steady enhancement in $V$ for a given temperature with increasing water concentration in both sides of the critical concentration. {\bf Inset:} The spin probe effective volume as a function of temperature for different glycerol concentrations. Solid lines are drawn through the points for clarity.}
\end{figure}

In the context of our discussion, this effective volume is also a measure of the available free volume of the host and thus can provide microstructural insights of the matrix. Fig \ref{fig5} represents the calculated effective volumes as a function of concentration at three isotherms 303K, 293 K and 283 K. A remarkable change in the quantity is seen in the vicinity of the critical glycerol concentration of 40 mol\%. The calculated volumes at different concentrations, which are also a measure of the DSE ratio $\mathcal R_{DSE} =\eta/(\tau T)$, are found to be only weakly temperature dependent (Fig \ref{fig5}, inset). This validates our statement about the strong coupling of spin probe dynamics with the viscosity of G-W system in the temperature range 283 - 303 K. Noteworthy in this context, a recent room temperature X-ray diffraction study on G-W system showed sudden change in the average intermolecular distance $(L_{m})$ in the vicinity of 40 mol\% glycerol content \cite{Hayashi:2007}.
The addition of smaller molecules like water help to open up void spaces in the glycerol matrix enhancing the available free volume of the system, albeit the average intermolecular distance of the system decreases. In a mixture with more water content that the critical concentration of 40 mol\%, the formation of water cooperative domain can bring about non-linear enhancement of the system's free volume which induces a sudden upswing in $V$ as presented in fig \ref{fig5}. Consistent with this view, the drop in average intermolecular distance $L_{m}$ across the critical concentration \cite{Hayashi:2007} is due to the presence of smaller water molecules {\it in excess}.

The strong correlation between the fragility of a glassformer and its free volume can be used to establish the reliability of our finding. The glassformer with higher degree of fragility usually possess more free volume \cite{shiro:97}. As we have already discussed, the fragility of G-W mixture enhances on addition of water. So, for a given temperature the enhancement of spin probe effective volume on each side of the critical concentration with increasing water content as seen in fig \ref{fig5} can be directly correlated with fig \ref{fig4} which demonstrate the simultaneous increase in fragility of the system. This proves the strong correlation of the spin probe effective volume $V$ with the system free volume.

A recent study of the Stokes-Einstein (SE) relation, stating the constance of the quantity $\mathcal R_{SE} = D\eta /T$ ($D$ is the translational diffusion coefficient), showed that the onset temperature of SE breakdown in supercooled G-W mixture decreases with increasing water content \cite{Sigmund:2006}. The SE relation is found to be closely obeyed for 85 mol\% ($\sim$ 96 wt\%) glycerol down to $\simeq$ 280 K. So, the SE relation is expected to hold at even lower temperatures for lower glycerol contents. Theoretical analysis \cite{Tarjus:1995} supported by recent simulations \cite{Poole:2006, Kob:2009} suggested that the SE and DSE laws follow the same molecular origin and violation of one leads to the violation of the other. In fact, the rotational diffusion of a system remain coupled to the shear viscosity down to much lower temperature than the translational diffusion \cite{Kob:2009}. This further validates our effective volume calculations in the temperature range of 283 - 303 K assuming the validity of DSE law where the SE law holds well for G-W mixtures. Moreover, the increase in the hydrodynamic radius with lowering of glycerol concentration, as observed in this study \cite{Sigmund:2006} substantiates the enhancement of free volume with increasing water content of the G-W system (Fig. \ref{fig5}).

\section{Conclusion}
A systematic study of the glass transition of G-W mixture has been carried out over the concentration range of 5 - 85 mol\% by the method of spin probe ESR. The estimated $T_{g}$\,s follow a concentration dependence which is closely associated to the mesoscopic inhomogeneities of the system. Formation of large scale inhomogeneities in water rich mixture is confirmed by effective volume estimation of the tracer within a temperature range of 283 - 303 K where tracer dynamics is strongly coupled to the system viscosity.

\end {document}